\documentclass{iopart}

\usepackage{graphicx}
\usepackage{mathrsfs}
\usepackage{hyperref}
\usepackage{geometry}
\usepackage{xcolor}
\usepackage{mathptmx}
\newcommand{\text}[1]{\mathrm{#1}}

\begin{document}

\title{Skyrmion behavior in attractive-repulsive square array of pinning centers}

\author{L. Basseto$^1$, N. P. Vizarim$^2$, 
J. C. Bellizotti Souza$^2$ and P. A. Venegas$^1$}

\address{$^1$ Department of Physics, Schools of Sciences, São Paulo State University - UNESP, 17033-360 Bauru, SP, Brazil}
\address{$^2$ "Gleb Wataghin" Institute of Physics, Department of Condensed Matter Physics, University of Campinas - UNICAMP, 13083-859, Campinas, SP, Brazil}

\ead{lucas.basseto@unesp.br}

\begin{abstract}
We investigate the driven dynamics of a single skyrmion in a square lattice of mixed pinning sites, where attractive and repulsive defects coexist using a particle-based model. The mixed landscape yields directional locking at $\theta_{\rm sk}=-45^\circ$ and flow at locked angles near the intrinsic skyrmion Hall angle. By mapping defect strengths, we show that weaker attraction lowers the depinning threshold, whereas stronger repulsion stabilizes and broadens the $-45^\circ$ locking plateau. Moreover, combinations of attractive and repulsive defect strengths allows control of directional lockings and their force ranges. Defect size further tunes the response, selecting among $-45^\circ$, $-50^\circ$, $-55^\circ$, and $\approx-59^\circ$. These results establish mixed pinning as a practical knob to steer skyrmion trajectories and the effective Hall response, providing design guidelines for skyrmion-based memory and logic devices.
\end{abstract}


\maketitle

\section{Introduction}
Magnetic skyrmions are topological excitations stabilized in chiral magnets by the interplay between exchange and the Dzyaloshinskii–Moriya interaction (DMI) \cite{bogdanov1994vortex, bogdanov1989vortices}. Owing to their nontrivial topology, skyrmions are robust against moderate perturbations and can be treated as quasiparticles. Beyond their fundamental interest, skyrmions are promising candidates for information carriers in spintronic applications, where the bit state can be encoded in their presence or absence \cite{kiselev2011,hagemeister2015,luo_reconfigurable_2018,zhang_magnetic_2015}, and they have also been proposed as building blocks for qubits and novel logic elements \cite{PhysRevLett.130.106701,romming2013writing}.

A central challenge in such skyrmion-based devices is that skyrmion motion is strongly influenced by a Magnus force set by topology \cite{zang_dynamics_2011,everschor-sitte_real-space_2014}. In clean samples, a driven skyrmion travels at an intrinsic angle $\theta_{\rm sk}^{\rm int}$ relative to the drive, a phenomenon known as the skyrmion Hall effect (SkHE) \cite{jiang_direct_2017,litzius_skyrmion_2017}. The magnitude of $\theta_{\rm sk}^{\rm int}$ increases with the ratio of the Magnus term to the dissipative term, and experimentally spans from a few degrees to very large values depending on material parameters and skyrmion size \cite{jiang_direct_2017,zeissler_diameter-independent_2020,litzius_skyrmion_2017}. Large deflections are detrimental for racetrack-like devices, since skyrmions may leave the track and be annihilated at the edges \cite{zhang_magnetic_2017,fert_magnetic_2017}.

In realistic media, pinning strongly impacts trajectories and effectively changes the observed skyrmion Hall angle $\theta_{\rm sk}$ \cite{reichhardt_collective_2015,reichhardt2022skyrmions}. These pinning centers can be either attractive or repulsive, natural or nanoengineered. Remarkably, periodic pinning has been shown to quantize or lock the direction of motion as the drive increases, enabling controllable transport \cite{reichhardt_quantized_2015,vizarim_skyrmion_2020,vizarim_skyrmion_2020-1,feilhauer_controlled_2020,vizarim_directional_2021,stosic_pinning_2017}. This directional locking is analogous to phenomena in superconducting vortices \cite{reichhardt_phase_1999} and colloids \cite{bohlein_experimental_2012,reichhardt_directional_2004,gopinathan_statistically_2004} driven over periodic substrates. For skyrmions, the drive direction is fixed, but the velocity-dependent $\theta_{\rm sk}$ produces discrete reorientations of the trajectory as the drive magnitude is ramped. On a locking step, $\theta_{\rm sk}$ remains constant over a finite drive interval, while transitions between steps coincide with characteristic features (dips or cusps) in the velocity–force curves. Such plateaus offer a practical knob to steer skyrmions via small drive adjustments. Additional strategies to control motion include ratchet effects \cite{gobel_skyrmion_2021,ma_reversible_2017,reichhardt_magnus-induced_2015,souza_skyrmion_2021,vizarim_skyrmion_2020-2,souza_controlled_2024}, interface-guided transport \cite{vizarim_guided_2021,zhang_edge-guided_2022}, strain \cite{yanes_skyrmion_2019}, magnetic or temperature gradient drives \cite{zhang_manipulation_2018,casiraghi_individual_2019,everschor_rotating_2012,kong_dynamics_2013,wang_rectilinear_2021}, 1D potential channels \cite{purnama_guided_2015,juge_helium_2021,kern_deterministic_2022}, nanotracks \cite{leliaert_coupling_2018,zhang_magnetic_2015,toscano_suppression_2020}, and hybrid skyrmion–vortex architectures \cite{menezes_manipulation_2019,neto_mesoscale_2022,xie_visualization_2024}.

For square arrays of periodic obstacles, the locked directions often follow
$\phi=\arctan(p/q)$ with integers $p$ and $q$ \cite{reichhardt_quantized_2015,vizarim_skyrmion_2020}. However, to the best of our knowledge, the impact of coexisting attractive and repulsive sites (mixed sites) arranged on a square lattice has not been systematically explored on the skyrmion dynamics.

In this work we address this gap by studying skyrmion motion in a lattice combining attractive and repulsive defects. We demonstrate new ways for steering trajectories and tuning the effective Hall response, thereby refining control of skyrmion transport. Our simulation methodology (shown in Section \ref{simulation}) employs a particle-based approach for a single skyrmion \cite{Lin_2013}, with dissipative and Magnus terms in the equation of motion. Defect interactions are modeled by Gaussian potentials, and an external drive $F_D$ is applied. We characterize the direction of skyrmion motion by the Hall angle $\theta_{\rm sk}=\arctan(v_\perp/v_\parallel)$, defined from velocity components perpendicular and parallel to $F_D$.
Section \ref{dynamics} we present our results where three dynamical regimes can be seen: (i) a pre-depinning regime where the skyrmion is trapped; (ii) a directional-locking regime with a stable trajectory at $-45^\circ$; and (iii) a high-drive Magnus regime where $\theta_{\rm sk}\!\to\!\theta_{\rm sk}^{\rm int}$. The novelty here is that the relative strengths of attractive and repulsive pinning control the extent of each regime, and that sufficiently strong attraction suppresses the locking plateau. In Section \ref{strength} we map the pinning strength effects to identify optimal operating windows. In Section \ref{size} we investigate pinning size effects and find that enlarging repulsive sites can modify $\theta_{\rm sk}$ at low drives, while at higher drives the motion is dominated by the intrinsic SkHE and becomes less sensitive to size. In Section \ref{discussion} we discuss our results highlighting important features and addressing future prospects, and in Section \ref{summary} a summary is exhibited.

\section{Simulation}
\label{simulation}

\begin{figure}
    \centering
    \includegraphics[width=0.5\textwidth]{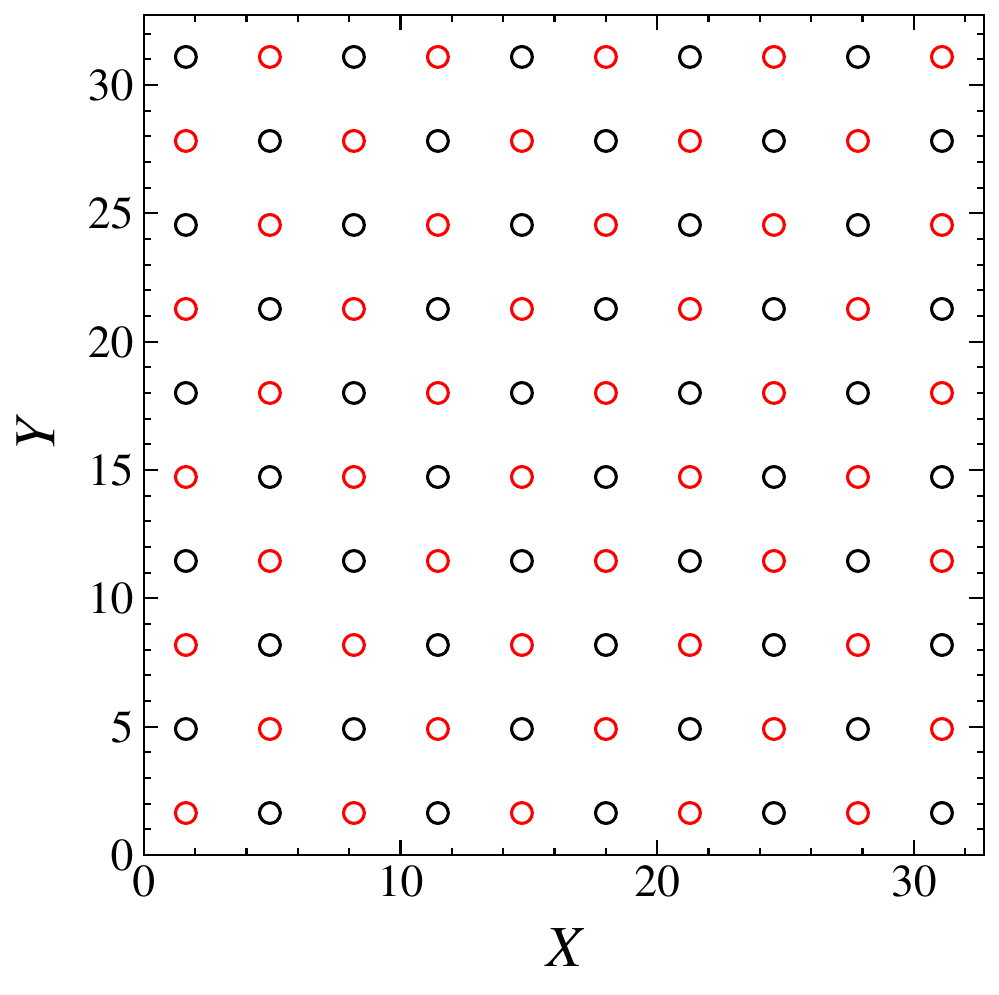}
    \caption{Schematic of the square array of periodic defects in our system that interact with the skyrmion. Red circles correspond to repulsive obstacles, while black circles to attractive pinning sites.}
    \label{fig:rede_quadrad}
\end{figure}

We simulated the dynamics of a single skyrmion in a ferromagnetic thin film that can host N\'eel skyrmions at zero temperature. The skyrmion is embedded in a two-dimensional simulation box of dimensions $L_x \times L_y$, with periodic boundary conditions along the $x$ and $y$ directions. As shown in Fig.~\ref{fig:rede_quadrad}, the system contains a square lattice of nano\-engineered defects that are either attractive or repulsive to skyrmions.

The skyrmion dynamics follows a particle-based equation of motion as in Lin \textit{et al.} \cite{Lin_2013}. We use a custom Fortran code based on standard molecular dynamics methods \cite{Molecular_book}. The equation of motion is:

\begin{equation}
\alpha_d \mathbf{v}_i + \alpha_m \, \hat{z} \times \mathbf{v}_i = 
\mathbf{F}_i^{\text{at}} + \mathbf{F}_i^{\text{rep}} + \mathbf{F}_D
\end{equation}

The first term on the left-hand side represents the damping that arises from 
spin precession and electron dissipation in the skyrmion core,
where $\alpha_d$ is the damping constant. 
The second term represents the Magnus force, acting 
perpendicularly to the skyrmion velocity, where $\alpha_m$ is the Magnus constant.
On the right side of the equation we have 
the skyrmion-defect interactions, $\mathbf{F}_i^{\text{at}}$ and $\mathbf{F}_i^{\text{rep}}$, that are modeled by a Gaussian potential of the type 
$ U_o = C_0 \, e^{-(r_{io}/a_o)^2} $, where $ C_0 $ represents the strength 
of the defect potential. The force associated with this potential is 
$ \mathbf{F}_{io} = -\nabla U_o = -\frac{2U_o}{a_o^2} \, r_{io} \, e^{-(r_{io}/a_o)^2} \, \hat{r}_{io} $. Here, $ r_{io} $ is the distance between 
skyrmion $ i $ and pinning site $ o $, $ a_o $ is the pinning radius, 
and $ \hat{r}_{io} $ is the unit vector pointing from the pinning center to the skyrmion.
As mentioned earlier, in this work two types of obstacles were considered.
Attractive pinning sites are described 
by a negative potential strength $C_0 = - C_{\rm at}$, 
producing a force that pulls the skyrmion toward the defect center. 
While repulsive obstacles have 
a positive potential strength $C_0 = C_{\rm rep}$ 
resulting in a force that pushes the skyrmion away from the obstacle center.
Thus, each pinning or obstacle site have different potential strengths $C_{\rm at}$ and $C_{\rm rep}$, 
and different radius $a_{\rm at}$ and $a_{\rm rep}$, respectively.
The third term on the right side is the interaction between the skyrmion
and the external transport force, given by $ \mathbf{F}_D = F_D \, \hat{d} $, 
where throughout this work we use $ \hat{d}  = \hat{x}$.
We increase $F^{D}$ in small steps of $\delta F^{D}= 0.01$
and spend $2\times 10^{5}$
simulation time steps, with a time step $\delta t=0.001$, at each drive increment to ensure a steady state.

We measure the 
average velocities $\left\langle V_\parallel\right\rangle=\left\langle \mathbf{v} \cdot \widehat{\rm 
{\bf{x}}}\right\rangle$ and $\left\langle V_\perp\right\rangle = \left\langle \mathbf{v} \cdot
\widehat{\rm {\bf{y}}}\right\rangle$.
We normalize all distances by
the screening length $\xi$ and select
the damping and Magnus constants
such that ${\alpha_m}^2+{\alpha_d}^2=1$.
Besides that, we also fix the rate $(\alpha_m / \alpha_d)= 1.732$, resulting in an intrinsic 
Hall angle of approximately $ \theta_{\text{sk}} \approx -60^\circ $ in all calculations.

\section{Skyrmion Dynamics with Attractive and Repulsive Defects}
\label{dynamics}

\begin{figure}
    \centering
    \includegraphics[width=0.7\textwidth]{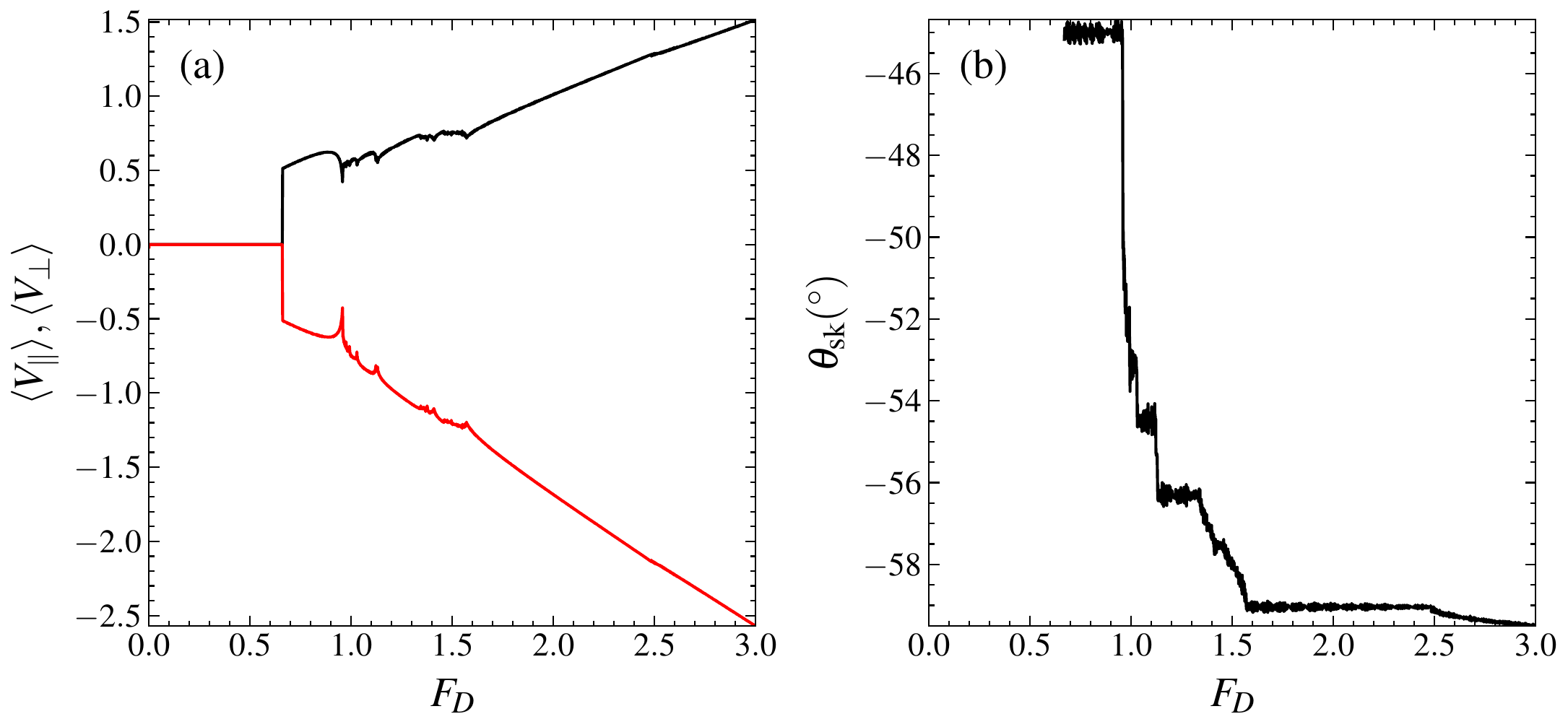}
    \caption{Results for a system with a single skyrmion in mixture of repulsive and attractive pinning sites as shown in Fig. \ref{fig:rede_quadrad}, using $a_{at} = a_{rep}=0.65$, $C_{\rm rep}=-0.5$ and $C_{\rm rep}=1.0$.
    (a) $\langle V_{\parallel} \rangle$ and $\langle V_{\perp} \rangle$ as a function $F_D$. 
(b) The corresponding Hall angle $\theta_{sk}$ as a function of $F_D$.}
    \label{fig:vel_fatr0.5_frep1}
\end{figure}

\begin{figure}
    \centering
    \includegraphics[width=0.7\textwidth]{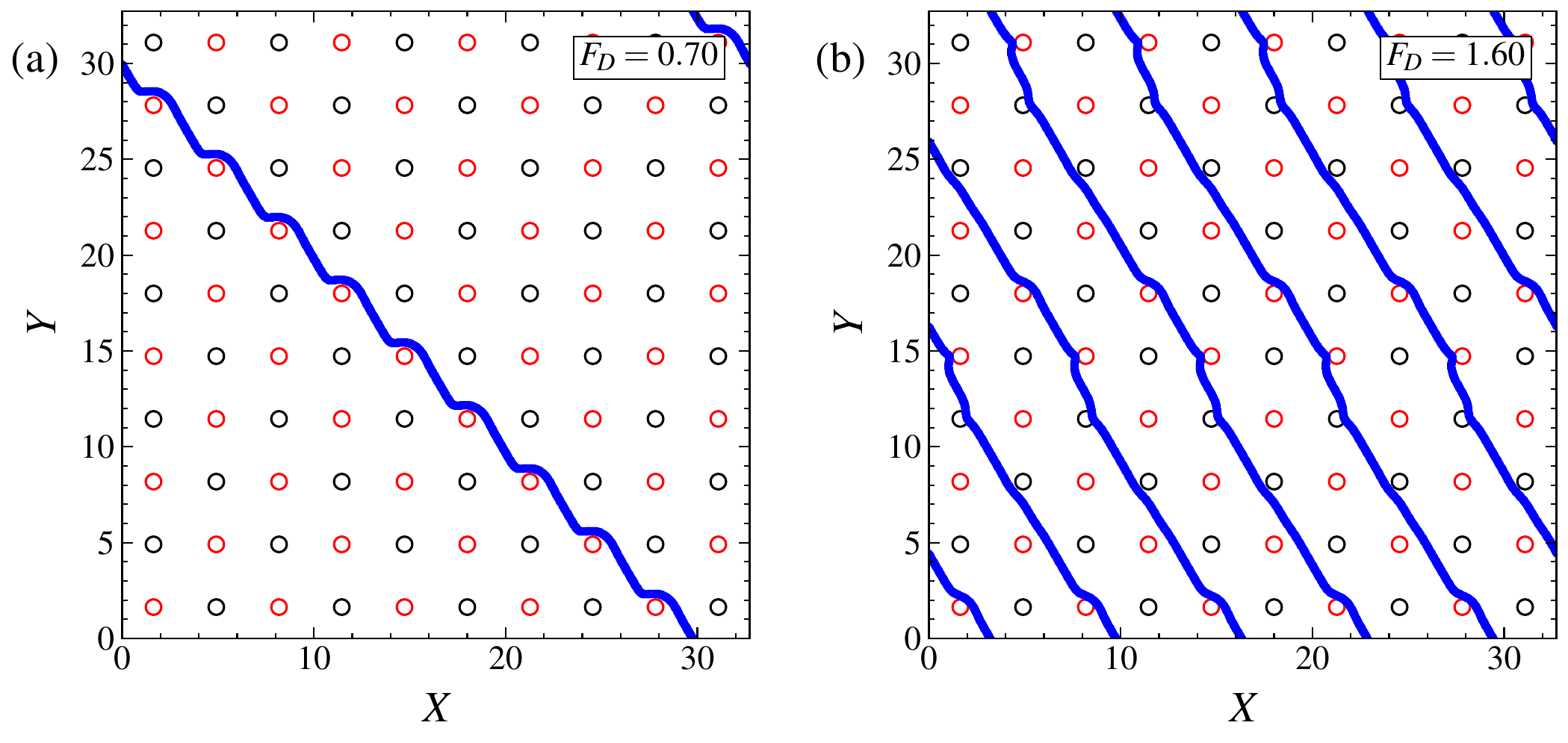}
    \caption{Illustration of the skyrmion trajectory (blue line) under the defect array for different values of the external driving force $F_D$ for the system shown in Figs. \ref{fig:rede_quadrad}. and \ref{fig:vel_fatr0.5_frep1}. In (a) $F_D = 0.70$, where a directional locking regime occurs with $\theta_{\rm sk}=-45^{\circ}$; (b) $F_D = 1.60$, regime dominated by a directional locking $\theta_{\rm sk}=-59^{\circ}$ very close to the intrinsic Hall angle $\theta_{\rm sk}^{\rm int}\approx-60^{\circ}$.}
    \label{fig:trajetoria}
\end{figure}

We first consider $C_{\rm at}=-0.5$, $C_{\rm rep}=1.0$, and $a_{\rm at}=a_{\rm rep}=0.65$. Fig.~\ref{fig:vel_fatr0.5_frep1} shows $\langle V_\parallel\rangle$, $\langle V_\perp\rangle$ and the corresponding $\theta_{\rm sk}$ as functions of $F_D$.
The initial skyrmion position, $i.e.$, the ground state is the skyrmion trapped in a attractive pinning site close to the center of the simulation box. It is important to note that the energy of the system is the same for the skyrmion trapped at any attractive pinning site since we consider an infinite ferromagnetic thin-film with periodic boundary conditions. As we increase the external driving, $F_D$, the skyrmion average velocities remain with zero values for $F_D \leq 0.68$, as shown in Fig. \ref{fig:vel_fatr0.5_frep1} (a). This means that the skyrmion remain trapped in the attractive defect, and its equilibrium position just deviates a bit from the center of the defect due to the external driving. This is the \textit{pinned phase}. For $F_D > 0.68$, Fig. \ref{fig:vel_fatr0.5_frep1} (a) shows that  $\left\langle V_\parallel\right\rangle$ exhibit a sharp increase and $\left\langle V_\perp\right\rangle$ a sharp decrease, indicating that the skyrmion begin to move. For the interval $0.68 < F_D < 0.95$ the average velocities varies smoothly with the applied drive. This regime can be seen in Fig. \ref{fig:vel_fatr0.5_frep1} (b) where skyrmion direction of motion is locked at $\theta_{\rm sk} = 45^{\circ}$. This is a clear \textit{directional locking} effect for skyrmions in a periodic array of repulsive obstacles that has been observed in previous works \cite{reichhardt_quantized_2015,vizarim_skyrmion_2020,feilhauer_controlled_2020}. However, here we show that this phase is also observed in a mixture between attractive and repulsive defects. Due to the ratio $\alpha_m/\alpha_d = 1.732$, the skyrmion intrinsic Hall angle is $\theta_{\rm sk}^{\rm int}\approx 60^{\circ}$. However, the interaction between the pinning landscape, the skyrmion and the external drive results in a directional locking at $\theta_{\rm sk}=45^{\circ}$. Different than observed in previous works \cite{reichhardt_quantized_2015,vizarim_skyrmion_2020,feilhauer_controlled_2020}, where only repulsive obstacles were considered, here we do not observe the skyrmion moving at $\theta_{\rm sk}=0^{\circ}$. In order for the skyrmion to move along $\theta_{\rm sk}=0^{\circ}$ it is necessary close repulsive defects to repel the skyrmion and lock its motion along the driving force. Here, due to the mixture of repulsive and attractive defects, we observed a expanded range of forces of $\theta_{\rm sk}=45^{\circ}$.
In Fig. \ref{fig:trajetoria} (a) we illustrate the skyrmion trajectories among the defects. Note that the skyrmion moves along the diagonal of the sample, being repelled by the repulsive obstacles in a zig-zag type of motion.

For $F_D > 0.95$, both skyrmion average velocities, $\left\langle V_\parallel\right\rangle$ and $\left\langle V_\perp\right\rangle$, exhibit decrease in magnitude followed by oscillations, as shown in Fig. \ref{fig:vel_fatr0.5_frep1} (a), indicating the presence of a transient skyrmion motion between stable phases. This can also be seen in Fig \ref{fig:vel_fatr0.5_frep1} (b), where $\theta_{\rm sk}$ is oscillating and changing locking directions in short intervals of $F_D$. For $1.13 < F_D < 1.42$, the skyrmion velocities increase in magnitude more smoothly, along with a stable $\theta_{\rm sk}\approx 59^{\circ}$. For  $F_D > 2.50$,  the skyrmion direction of motion tends to align with the intrinsinc skyrmion Hall angle continuously, similar to observed in previous works \cite{reichhardt_quantized_2015,vizarim_skyrmion_2020,vizarim_directional_2021}.
This is a clear signature that pinning effects are being mitigated by the strong driving forces, resulting in a dynamics that resembles the clean sample case.
In Fig. \ref{fig:trajetoria} (b) it is illustrated some skyrmion trajectories for $F_D = 1.60$, where the system is locked along $\theta_{\rm sk}\approx59^{\circ}$. In this case, the skyrmion trajectory passes through some attractive pinning sites but also some repulsive sites, different than shown in Fig. \ref{fig:trajetoria} (a).

\section{Effects of Defect Strength on the Dynamics}
\label{strength}

\begin{figure}
    \centering
    \includegraphics[width=0.7\textwidth]{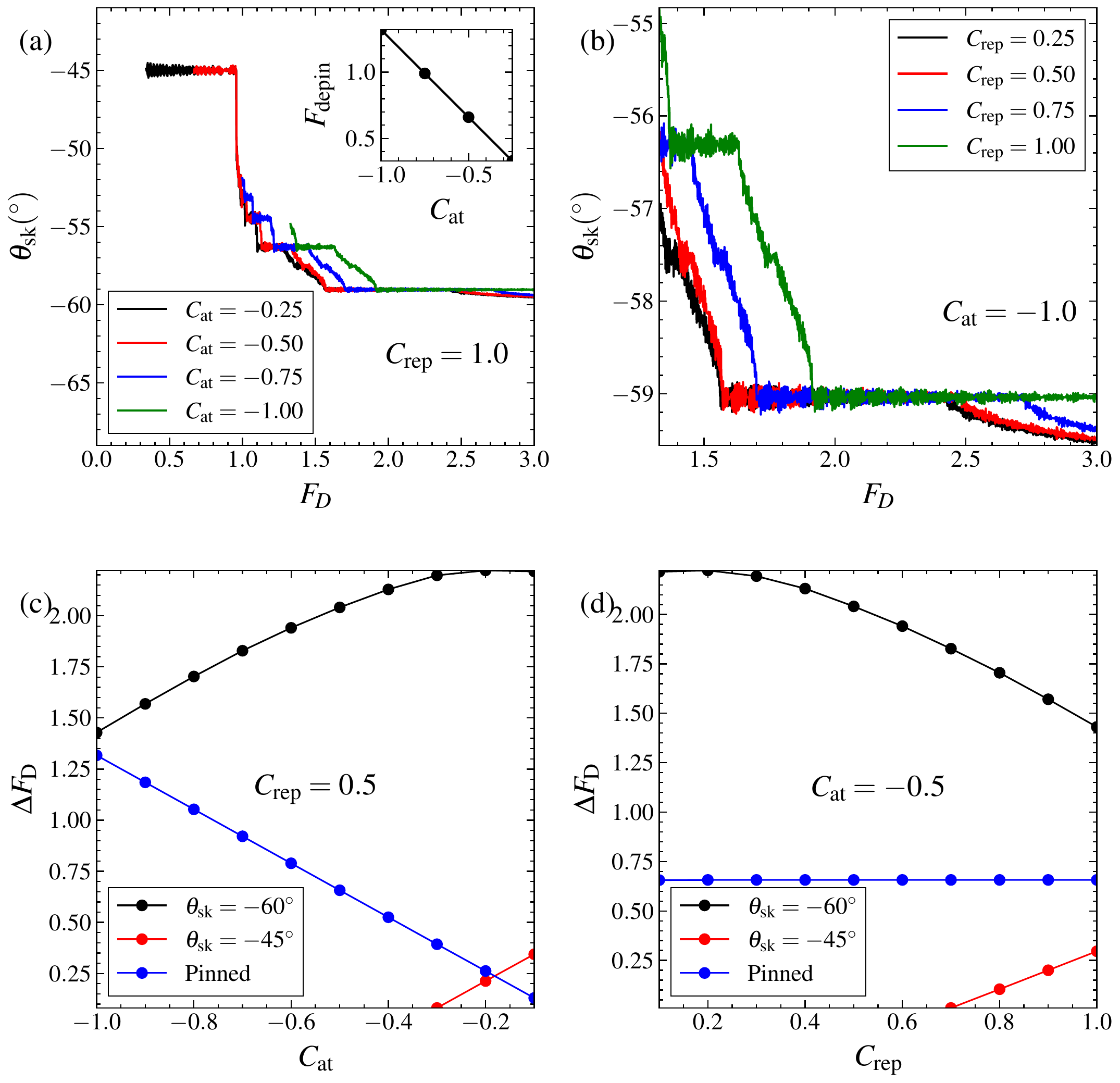}
    \caption{ (a,b) Skyrmion Hall angle $\theta_{\rm sk}$ as a function of the driving force, $F_D$ and (c,d) the range of forces $\Delta F_D$ vs. the defect strengths.
    (a) Results for fixed $C_{\rm rep}= 1.0$ and varied $C_{\rm at}$, and (b) for fixed $C_{\rm at}= -1.0$ and varied $C_{\rm rep}$. (c) $\Delta F_D$ vs. $C_{\rm at}$ with fixed $C_{\rm rep}= 0.5$ and (d) $\Delta F_D$ vs. $C_{\rm rep}$ with $C_{\rm at}= -0.5$. Simulations were performed with the ratio $\alpha_m/\alpha_d = 1.732$ and $a_{\rm at}=a_{\rm rep} = 0.65$.}
    \label{fig:Fat_Frep_theta}
\end{figure}

In Fig.~\ref{fig:Fat_Frep_theta} (a), we plot $\theta_{\rm sk}$ as a function of the driving force $F_D$ for a system with both defect radius fixed at $a_{\rm at} = a_{\rm rep} =0.65$. The repulsive strength was kept constant at $C_{\rm rep} = 1.0$, while the attractive strength $C_{\rm at}$ was varied using $-0.25$, $-0.50$, $-0.75$, and $-1.0$. The results clearly show that the depinning threshold is sensitive to the magnitude of the attractive strength $C_{\text{at}}$. In the inset of Fig.~\ref{fig:Fat_Frep_theta} (a), it is shown a clear linear increase in magnitude of $F_{\rm depin}$ as a function of $C_{\rm at}$. Additionally, we observe that the regime characterized by $\theta_{\text{sk}} = -45^\circ$ is dependent of $C_{\text{at}}$. For example, the range of $F_D$ in which the skyrmion remains in this regime is largest for $C_{\text{at}} = -0.25$. On the other hand, this regime vanishes completely for $|C_{\text{at}}| \geq 0.75$. That is, for stronger values of $C_{\rm at}$ the skyrmion becomes strongly pinned at the pinning site. When the external drive is sufficient large to depin the skyrmion, it depins with high velocity and cannot stabilize the regime at which it locks at $\theta_{\rm sk}=-45^{\circ}$. This $\theta_{\rm sk}=-45^{\circ}$ locking phase requires moderate skyrmion velocities to be stable.

The results for  $\theta_{\rm sk}$ vs. $F_D$ for the system with fixed $a_{\rm at} = a_{\rm rep} =0.65$ and $C_{\rm at} = -1.00$, while varied $C_{\rm rep} = 0.25$, $0.50$, $0.75$, and $1.0$ is shown in Fig. \ref{fig:Fat_Frep_theta} (b). Different than shown in Fig. \ref{fig:Fat_Frep_theta} (a), here the depinning thresholds are all the same, at $F_D=1.34$. Since the ground state correspond to a skyrmion trapped at an attractive pinning site, the effect of $C_{\rm rep}$ has no effect on the depinning forces. However, the dynamic phases strongly depends on $C_{\rm rep}$. For the case of $C_{\rm rep} = 0.25$ the skyrmion direction of motion tends to the intrinsic angle much faster than for stronger values of $C_{\rm rep}$. That is, the repulsive obstacle sites are crucial for controlling the skyrmion direction of motion, corroborating previous works \cite{vizarim_skyrmion_2020}. For example, using $C_{\rm rep}=1.0$, the intervals at which the dynamic phases are stable are much expanded, including the $\theta_{\rm sk} \approx 59^{\circ}$ phase.
These features can also be seen in Fig. \ref{fig:Fat_Frep_theta} (c) and (d), where we performed a detailed analysis of the range of driving force regimes $\Delta F_D$ for selected most prominent Hall angles, $\theta_{\rm sk}=-45^{\circ}$ and $-60^{\circ}$. These ranges of $\Delta F_D$ are determined directly from the $\theta_{\rm sk}$ vs $F_D$ curves shown in (a) and (b), identifying the force ranges in which the $\theta_{\rm sk}$ remains fairly constant within a tolerance of $\Delta \theta_{\rm sk}= \pm1.5 ^{\circ}$ due to numeric oscillations. Fig. \ref{fig:Fat_Frep_theta} (c) illustrates the intervals $\Delta F_D$ for the pinned state, $\theta_{\rm sk}=-45^\circ$ and $-60^\circ$ as a function of the attractive force $C_{\rm at}$, while fixed $C_{\rm rep} = 0.5$ for several values of $C_{\rm at}$. From Fig. \ref{fig:Fat_Frep_theta} (c), it is possible to see that the pinned phase progressively decreases as $|C_{\rm at}|$ decreases. As the attractive sites becomes weaker, it is easier for the skyrmion to depin. The $\theta_{\rm sk}=-45^\circ$ regime can only be stable for $|C_{\rm at}| \leq 0.3$. Additionally, the interval of forces $\Delta F_D$ where this regime is stable increases with decreasing $C_{\rm at}$. The phase where the regime is locked at $\theta_{\rm sk}=-60^\circ$ also increases as $|C_{\rm at}|$ decreases. This result is expected, as the attractive pinning sites becomes weaker, the system tends to behave more similar to a clean sample where the stable regime is flowing along the intrinsic angle. In Fig. \ref{fig:Fat_Frep_theta} (d) is shown the intervals $\Delta F_D$ as a function of the repulsive force $C_{\rm rep}$, while fixed $C_{\rm at} = -0.5$. The pinned phase is completely independent of $C_{\rm rep}$. The $\theta_{\rm sk}=-45^\circ$ regime is stable for $C_{\rm rep} \geq 0.7$, while $\theta_{\rm sk}=-60^\circ$ phase tends to be more robust as $C_{\rm rep}$ decreases. These results show clearly that there is a balance of values of $C_{\rm rep}$ and $C_{\rm at}$ where more dynamic phases can be observed. Thus, from the practical point of view, it is possible to modulate phases using a combination of attractive and repulsive defect sites.

\begin{figure}
    \centering
    \includegraphics[width=0.7\textwidth]{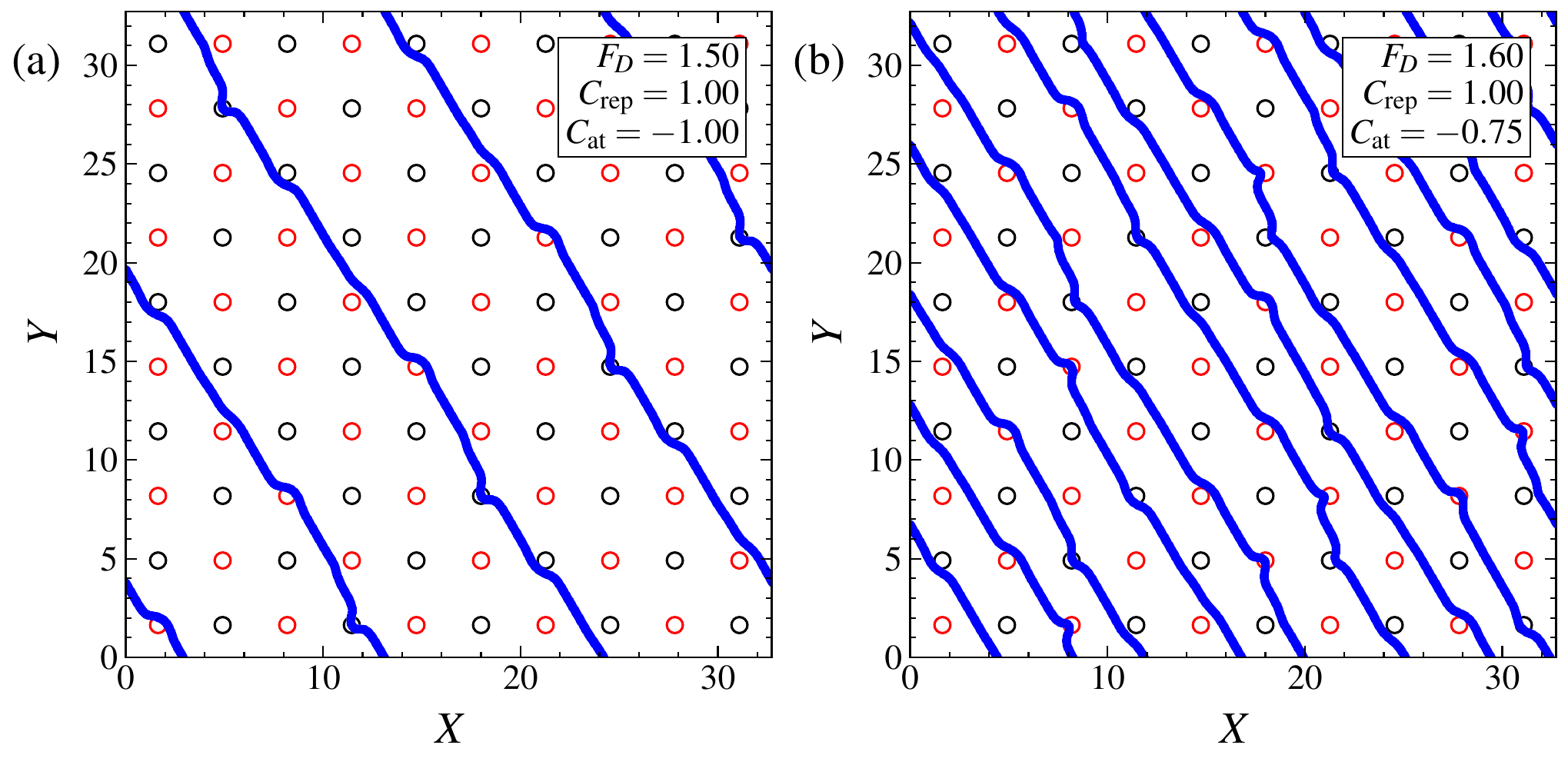}
    \caption{Skyrmion trajectories in a periodic lattice of attractive and repulsive defects for different combinations of potential strengths and driving force, $F_D$. 
    (a) $C_{\rm at} = -1.0$, $C_{\rm rep} = 1.0$ and $F_D=1.50$, where the skyrmion motion is locked at $\theta_{\rm sk}=-57^{\circ}$. 
    (b) $C_{\rm at}=-0.75$, $C_{\rm at}=1.0$ and $F_D=1.60$, where the skyrmion motion is in a transient state, between locked states $\theta_{\rm sk}=-57^{\circ}$ and $-59^{\circ}$. 
    \label{fig:Fat_trajetoria}}
\end{figure}

Fig. \ref{fig:Fat_trajetoria} illustrates selected skyrmion trajectories.
In Fig. \ref{fig:Fat_trajetoria} (a), the skyrmion is flowing in a defect landscape using $C_{\rm at} = -1.0$, $C_{\rm rep} = 1.0$ and $F_D=1.50$, where its motion is locked at $\theta_{\rm sk}\approx-57^{\circ}$. In this regime the skyrmion flows and interact with both attractive and repulsive sites resulting in a periodic motion displacing six lattice parameters along $y$ and four along $x$, that is, a ratio of $R=6/4$. The angle of motion can be calculated as $\theta_{\rm sk}=\text{arctan}(R) \approx 57^{\circ}$.
In transient Fig. \ref{fig:Fat_trajetoria} (b), using $C_{\rm at}=-0.75$, $C_{\rm at}=1.0$ and $F_D=1.60$, the skyrmion motion is not locked at a specific direction. This is a transient motion between different locking directions. As can be seen, the skyrmion motion does not exhibit periodicity.

\begin{figure}
    \makebox[\textwidth][c]{\includegraphics[width=0.9\textwidth]{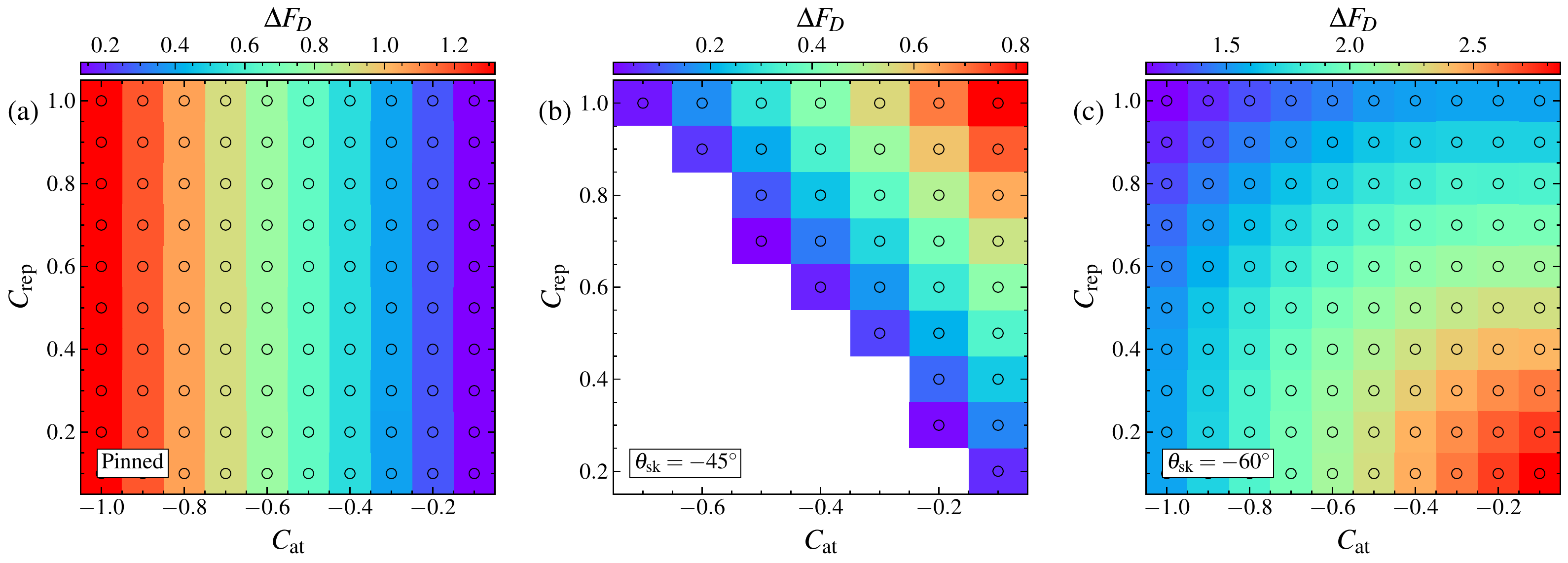}}
    \caption{Dynamic phases of $\Delta F_D$ as a function of attractive ($C_{\text{at}}$) and repulsive ($C_{\text{rep}}$) defect strengths using $\alpha_m/\alpha_d = 1.732$ and $a_{\rm at}=a_{\rm rep} = 0.65$. In (a) diagram for the pinned phase, (b) $\theta_{\text{sk}} = -45^\circ$; and (c) $\theta_{\text{sk}} = -60^\circ$.}
    \label{fig:diagrama}
\end{figure}

For a complete description of the defect strengths, $C_{\rm at}$ and $C_{\rm rep}$, in Fig. \ref{fig:diagrama} we show a phase diagram of $\Delta F_D$ vs. $C_{\rm at}$ vs. $C_{\rm rep}$ for the pinned phase, $\theta_{\rm sk}=-45^\circ$ and $\theta_{\rm sk}=-60^\circ$. In Fig. \ref{fig:diagrama} (a) it is clearly shown that the pinned phase depends solely on $C_{\rm at}$. In Fig. \ref{fig:diagrama} (b) we observe that $\theta_{\rm sk}=-45^\circ$ regime can only occur for a combination of low $C_{\rm at}$ and high $C_{\rm rep}$. The white region in the plot indicates regions where this phase cannot be stabilized. Finally, $\theta_{\rm sk}=-60^\circ$ is most prominent when both $C_{\rm rep}$ and $C_{\rm at}$ are weak. That is, when the system exhibits weak pinning, the dynamics resembles the clean sample case, where the skyrmion flows along the intrinsic SkHE, which in our case is $\theta_{\rm sk}^{\rm int} \approx 60^{\circ}$.

\section{Influence of Defect Size}
\label{size}

To investigate the effects of defect size on skyrmion dynamics, we fix $C_{\text{at}} = -0.2$ and $C_{\text{rep}} = 0.8$ and then vary the size of the attractive and repulsive defect sites. These values were chosen as they exhibit the phases $\theta_{\text{sk}} = -45^\circ$ and $ \approx -60^\circ$, allowing us to observe how only defect size effects the dynamic regimes. The radii of the defects, both attractive and repulsive, were varied from $0.05$ to $1.0$ in steps of $0.05$.  

\begin{figure}
    \centering
    \includegraphics[width=0.7\textwidth]{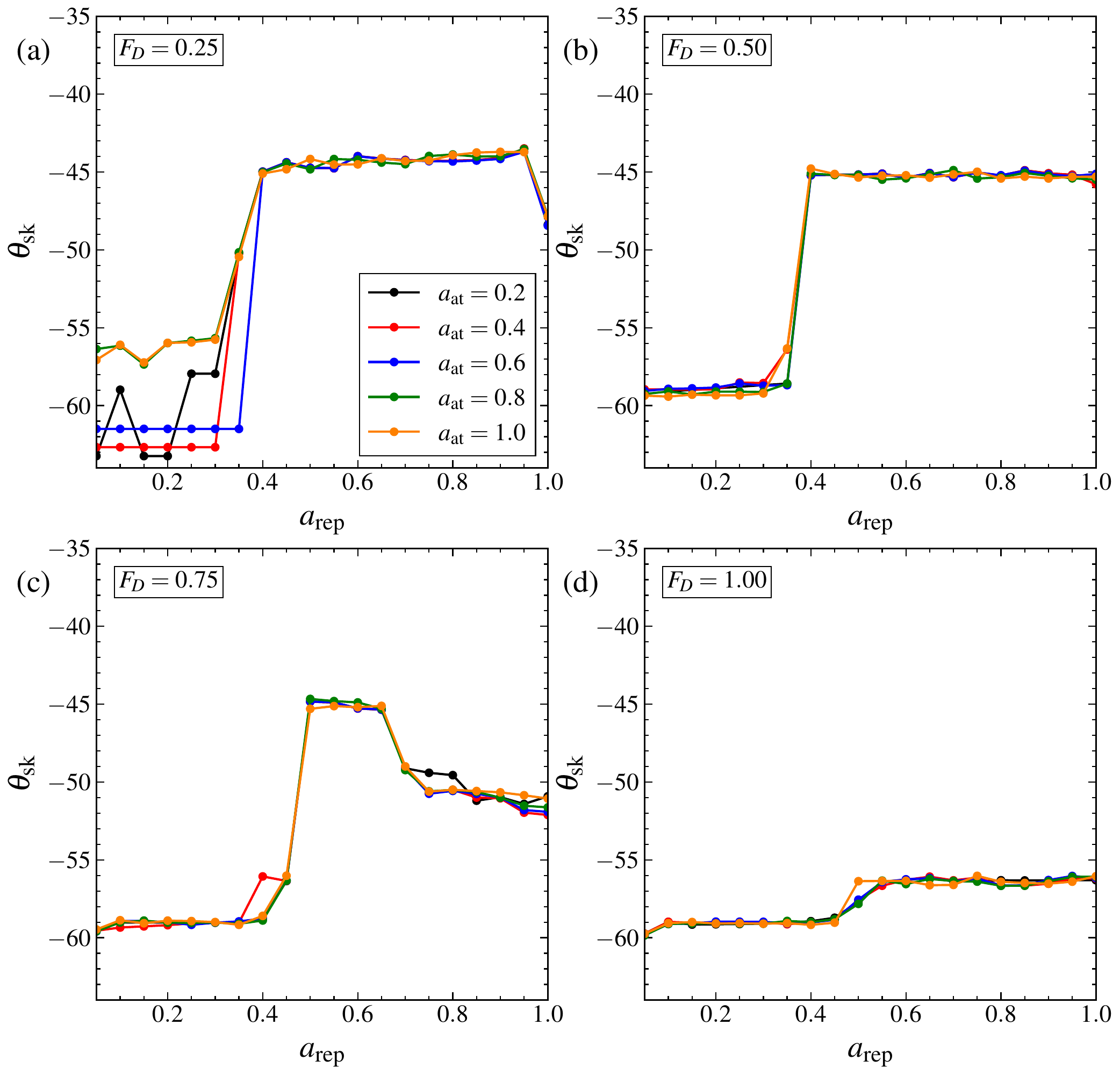}
    \caption{Skyrmion Hall angle, $\theta_{\rm sk}$ as a function of the repulsive defect size $a_{\text{rep}}$ for different attractive defect sizes $a_{\rm at}$ and driving force values: (a) $F_D = 0.25$, (b) $F_D = 0.50$, (c) $F_D = 0.75$, and (d) $F_D = 1.0$.}
    \label{fig:sk_fd}
\end{figure}

In Fig.~\ref{fig:sk_fd} (a), using $F_D = 0.25$, we can see that for small repulsive defects of $a_{\rm rep}<0.4$ the skyrmion Hall angle can lock at different angles $\theta_{\rm sk}\approx-56^{\circ}$ or $\approx-60^{\circ}$ depending on the value of $a_{\rm at}$. For small $a_{\rm at}$ and $a_{\rm rep}$, the skyrmion tends to lock at angles very close to the intrinsic angle, as expected. However, for larger attractive sites, $a_{\rm at} \geq 0.8$, the influence of the defects is strong enough to lock skyrmions along $\theta_{\rm sk}\approx56^{\circ}$.
For larger repulsive defects of $0.40<a_{\text{rep}}<0.95$, the skyrmion is locked at $\theta_{\rm sk}\approx -45^\circ$ phase. For very large repulsive defects, $a_{\rm rep} = 1.0$ the skyrmion locks at a different direction $\theta_{\rm sk}\approx 48^{\circ}$.
In Fig. \ref{fig:sk_fd} (b), using $F_D= 0.50$, as we vary the size of both defects it is only possible to observe two different locking directions, $\theta_{\rm sk}\approx-45^{\circ}$ for $a_{\rm rep}>0.40$, or $\approx-60^{\circ}$ for $a_{\rm rep}<0.40$. In this case, as more external driving force is applied, the skyrmion has higher average velocities which helps it to lock at fewer directions.
For $F_D = 0.75$, as shown in Fig. \ref{fig:sk_fd} (c), the $\theta_{\rm sk}\approx-45^{\circ}$ is much reduced, only available at $0.50<a_{\rm rep}<0.65$. For $a_{\rm rep}>0.65$ the skyrmion direction of motion tends to align along $\theta_{\rm sk}\approx-50^{\circ}$. Meanwhile, for $F_D=1.00$, shown in Fig. \ref{fig:sk_fd} (d), the skyrmion is even less influenced by the defects, changing from $\theta_{\rm sk}\approx-60^{\circ}$ to $\approx-56^{\circ}$ due to larger $a_{\rm rep}$.

Hence, defect size is a key parameter for steering skyrmion trajectories. Increasing the radius amplifies pinning or repulsive forces and expand directional locking phases, whereas smaller defects weaken these effects. Additionally, we expect that large and strong defects stabilize more locking directions. These findings chart a practical route to controlling skyrmion motion via coexisting attractive and repulsive defects.

\section{Discussion}
\label{discussion}

Using particle-based simulations we showed that it is possible to control a single skyrmion direction of motion using a landscape composed of attractive and repulsive defects.
We model the skyrmions following the model proposed by Lin \textit{et al.} \cite{Lin_2013}, where skyrmions are treated as point-like rigid bodies that cannot deform or change size; however, in real materials, skyrmions may deform, change size or be annihilated. These features can change the dynamics or give rise to new phenomena, which could be explored further using continuum-based simulations. 
Thermal fluctuations, neglected here, are likewise expected to modify the behavior, where thermal creep at low drives can shift depinning thresholds and phase boundaries, and fragile plateaus of small $\Delta F_D$ values may disappear at moderate temperatures.
We expect the qualitative influence of mixed pinning to be robust across other pinning geometries, while the preferred locking directions reflect lattice symmetry. That is, in the square array considered here, the principal direction is $45^{\circ}$, whereas a triangular lattice should favor $30^{\circ}$ and $60^{\circ}$.

\section{Summary}
\label{summary}

In this work we have investigated the dynamics of a single skyrmion interacting with a square array of defects that coexists attractive and repulsive defects. The defect landscape produces directional locking at $\theta_{\rm sk}=-45^{\circ}$ and flow near the intrinsic skyrmion Hall angle, in our case $\theta_{\rm sk}\approx -60^{\circ}$. Unlike repulsive-only arrays, our system does not exhibit a $\theta_{\rm sk}=0^{\circ}$ step, as the alternation of pin types suppresses longitudinal locking.
By tuning defect strengths, we control both the depinning threshold and the extent of the $-45^{\circ}$ locking step, where weaker attraction lowers the depinning force, whereas stronger repulsion stabilizes directional locking and broadens its force window. Thus, the set and width of accessible locking steps can be expanded or suppressed by appropriate strength choices. We also investigated defect size effects. At low drives the dynamics is more sensitive to defect radii, enabling locking at $\theta_{\rm sk}=-45^{\circ}$, $-50^{\circ}$, $-55^{\circ}$, or $\approx -59^{\circ}$ depending on parameters; at higher drives the motion converges toward $\theta_{\rm sk}^{\rm int}$. We hope our results bring new insights in using a combined mixture of defects that are attractive and repulsive to enable new possibilities to control the skyrmion motion in future spintronic devices.

\ack

This work was supported by the Brazilian agencies Coordena\c{c}\~ao de Aperfei\c{c}oamento de Pessoal de N\'ivel Superior - CAPES, Conselho Nacional de Desenvolvimento Cient\'ifico e Tecnol\'ogico - CNPq, and Funda\c{c}\~{a}o de Amparo \`{a} Pesquisa do Estado de S\~{a}o Paulo - FAPESP.
L.B., N.P.V., and J.C.B.S acknowledge
funding from FAPESP (Grants: 2024/21139-1, 2024/13248-5, and 2022/14053-8, respectively).
We would like to thank FAPESP for providing the computational resources used in this
work (Grant: 2024/02941-1). 

\section*{References}
\bibliographystyle{iopart-num}
\bibliography{mybib}

\end{document}